\documentclass[10pt,a4paper,onecolumn]{article}
\usepackage{marginnote}
\usepackage{graphicx}
\usepackage[rgb, dvipsnames, svgnames, x11names]{xcolor}
\usepackage{authblk,etoolbox}
\usepackage{titlesec}
\usepackage{calc}
\usepackage{tikz}
\usepackage[unicode]{hyperref}
\hypersetup{%
    unicode=true,
    pdftitle={Phentax: A JAX implementation of the IMRPhenomTHM waveform
model},
    pdfauthor={Alessandro Santini, Diganta Bandopadhyay, Jonathan Gair},
    colorlinks=true,
    linkcolor=[rgb]{0.0, 0.5, 1.0},
    citecolor=Blue,
    urlcolor=[rgb]{0.0, 0.5, 1.0},
    breaklinks=true
}
\pdfcatalog{%
  /OutputIntents [
    <<
      /Type /OutputIntent
      /S /GTS_PDFA1
      /DestOutputProfile \the\pdflastobj\space 0 R
      /OutputConditionIdentifier (sRGB)
      /Info (sRGB)
    >>
  ]
}

\usepackage{caption}
\usepackage{orcidlink}
\usepackage{tcolorbox}
\usepackage{amssymb,amsmath}
\usepackage{ifxetex,ifluatex}
\usepackage{seqsplit}
\usepackage{xstring}

\usepackage{float}
\let\origfigure\figure
\let\endorigfigure\endfigure
\renewenvironment{figure}[1][2] {
    \expandafter\origfigure\expandafter[H]
} {
    \endorigfigure
}

\usepackage{fixltx2e} 

\NewDocumentCommand\citeproctext{}{}
\NewDocumentCommand\citeproc{mm}{%
  \begingroup\def\citeproctext{#2}\cite{#1}\endgroup}
\makeatletter
 \let\@cite@ofmt\@firstofone
 \def\@biblabel#1{}
 \def\@cite#1#2{{#1\if@tempswa , #2\fi}}
\makeatother
\newlength{\cslhangindent}
\newlength{\csllabelwidth}
\newenvironment{CSLReferences}[2] 
 {\begin{list}{}{%
  \setlength{\itemindent}{0pt}
  \setlength{\leftmargin}{0pt}
  \setlength{\parsep}{0pt}
  \ifodd #1
   \setlength{\leftmargin}{\cslhangindent}
   \setlength{\itemindent}{-1\cslhangindent}
  \fi
  \setlength{\itemsep}{#2\baselineskip}}}
 {\end{list}}
\usepackage{calc}

\usepackage[top=3.5cm, bottom=3cm, right=1.5cm, left=1.0cm,
            headheight=2.2cm, reversemp, includemp, marginparwidth=4.5cm]{geometry}

\titleformat{\section}
  {\normalfont\sffamily\Large\bfseries}
  {}{0pt}{}
\titleformat{\subsection}
  {\normalfont\sffamily\large\bfseries}
  {}{0pt}{}
\titleformat{\subsubsection}
  {\normalfont\sffamily\bfseries}
  {}{0pt}{}
\titleformat*{\paragraph}
  {\sffamily\normalsize}

\usepackage{fancyhdr}
\makeatletter
\let\ps@plain\ps@fancy
\fancyheadoffset[L]{4.5cm}
\fancyfootoffset[L]{4.5cm}

\definecolor{linky}{rgb}{0.0, 0.5, 1.0}

\newtcolorbox{repobox}
   {colback=red, colframe=red!75!black,
     boxrule=0.5pt, arc=2pt, left=6pt, right=6pt, top=3pt, bottom=3pt}

\newcommand{\ExternalLink}{%
   \tikz[x=1.2ex, y=1.2ex, baseline=-0.05ex]{%
       \begin{scope}[x=1ex, y=1ex]
           \clip (-0.1,-0.1)
               --++ (-0, 1.2)
               --++ (0.6, 0)
               --++ (0, -0.6)
               --++ (0.6, 0)
               --++ (0, -1);
           \path[draw,
               line width = 0.5,
               rounded corners=0.5]
               (0,0) rectangle (1,1);
       \end{scope}
       \path[draw, line width = 0.5] (0.5, 0.5)
           -- (1, 1);
       \path[draw, line width = 0.5] (0.6, 1)
           -- (1, 1) -- (1, 0.6);
       }
   }

\patchcmd{\@maketitle}{center}{flushleft}{}{}
\patchcmd{\@maketitle}{center}{flushleft}{}{}
\patchcmd{\@maketitle}{\LARGE}{\LARGE\sffamily}{}{}
\def\maketitle{{%
  
  \AB@maketitle}}
\renewcommand\AB@affilsepx{ \protect\Affilfont}
\renewcommand\AB@affilnote[1]{{\bfseries #1}\hspace{3pt}}
\renewcommand{\affil}[2][]%
   {\newaffiltrue\let\AB@blk@and\AB@pand
      \if\relax#1\relax\def\AB@note{\AB@thenote}\else\def\AB@note{#1}%
        \setcounter{Maxaffil}{0}\fi
        \begingroup
        \let\href=\href@Orig
        \let\protect\@unexpandable@protect
        \def\thanks{\protect\thanks}\def\footnote{\protect\footnote}%
        \@temptokena=\expandafter{\AB@authors}%
        {\def\\{\protect\\\protect\Affilfont}\xdef\AB@temp{#2}}%
         \xdef\AB@authors{\the\@temptokena\AB@las\AB@au@str
         \protect\\[\affilsep]\protect\Affilfont\AB@temp}%
         \gdef\AB@las{}\gdef\AB@au@str{}%
        {\def\\{, \ignorespaces}\xdef\AB@temp{#2}}%
        \@temptokena=\expandafter{\AB@affillist}%
        \xdef\AB@affillist{\the\@temptokena \AB@affilsep
          \AB@affilnote{\AB@note}\protect\Affilfont\AB@temp}%
      \endgroup
       \let\AB@affilsep\AB@affilsepx
}
\makeatother

\renewcommand\Affilfont{\sffamily\small\mdseries}
\ifnum 0\ifxetex 1\fi\ifluatex 1\fi=0 
  \usepackage[T1]{fontenc}
  \usepackage[utf8]{inputenc}

\else 
  \ifxetex
    \usepackage{mathspec}
    \usepackage{fontspec}

  \else
    \usepackage{fontspec}
  \fi
  \defaultfontfeatures{Scale=MatchLowercase}
  \defaultfontfeatures[\sffamily]{Ligatures=TeX}
  \defaultfontfeatures[\rmfamily]{Ligatures=TeX,Scale=1}

\fi
\IfFileExists{upquote.sty}{\usepackage{upquote}}{}
\IfFileExists{microtype.sty}{%
\usepackage{microtype}
\UseMicrotypeSet[protrusion]{basicmath} 
}{}

\PassOptionsToPackage{usenames,dvipsnames}{color} 
\ifLuaTeX
\usepackage[bidi=basic]{babel}
\else
\usepackage[bidi=default]{babel}
\fi
\babelprovide[main,import]{american}

\def\languageshorthands#1{}

\IfFileExists{parskip.sty}{%
\usepackage{parskip}
}{
\setlength{\parindent}{0pt}
\setlength{\parskip}{6pt plus 2pt minus 1pt}
}
\ifx\paragraph\undefined\else
\let\oldparagraph\paragraph
\renewcommand{\paragraph}[1]{\oldparagraph{#1}\mbox{}}
\fi
\ifx\subparagraph\undefined\else
\let\oldsubparagraph\subparagraph
\renewcommand{\subparagraph}[1]{\oldsubparagraph{#1}\mbox{}}
\fi
\usepackage{subcaption}
\ifLuaTeX
  \usepackage{selnolig}  
\fi

\title{Phentax: A JAX implementation of the IMRPhenomTHM waveform model}

\author[1%
\ensuremath\mathparagraph]{Alessandro Santini%
  \,\orcidlink{0000-0001-6936-8581}\,%
}
\author[1%
]{Diganta Bandopadhyay%
  \,\orcidlink{0000-0003-0975-5613}\,%
}
\author[1%
]{Jonathan Gair%
  \,\orcidlink{0000-0002-1671-3668}\,%
}

\affil[1]{Max Planck Institute for Gravitational Physics (Albert
Einstein Institute) Am Mühlenberg 1, 14476 Potsdam, Germany}
\affil[$\mathparagraph$]{Corresponding author}
\date{\vspace{-2.5ex}}

\begin{document}
\maketitle

\marginpar{

  \begin{flushleft}
  \sffamily\small

    {\bfseries Preprint:} Prepared for submission to Journal of Open
Source Software
  
  \vspace{2mm}
    {\bfseries Software}
  \begin{itemize}
    \setlength\itemsep{0em}
    \item \href{https://github.com/openjournals}{\color{linky}{Repository}} \ExternalLink
  \end{itemize}

  \vspace{2mm}

  \vspace{2mm}
  {\bfseries License}\\
  Authors of papers retain copyright and release the work under a Creative Commons Attribution 4.0 International License (\href{https://creativecommons.org/licenses/by/4.0/}{\color{linky}{CC BY 4.0}}).

  \end{flushleft}
}

\section{Summary}\label{summary}

The Laser Interferometer Space Antenna (LISA)
(\citeproc{ref-LISA:2024hlh}{Colpi \& others, 2024}) is a joint ESA-NASA
space-based gravitational-wave (GW) detector that is expected to be
launched in the mid-2030s. LISA will operate in the millihertz region of
the GW spectrum, opening a new window for observing gravitational waves
emitted by systems we cannot characterize with current ground-based
detectors. In particular, during its operational lifetime, LISA is
expected to be sensitive to thousands of white dwarf (WD) binaries in
our galaxy (\citeproc{ref-Korol:2020lpq}{Korol \& others, 2020}),
extreme mass ratio inspirals (EMRIs) of stellar-mass compact objects
into massive black holes (\citeproc{ref-Babak:2017tow}{Babak et al.,
2017}), stellar-mass black hole binaries (SOBHBs) in their early
inspiral phase (\citeproc{ref-Buscicchio:2021dph}{Buscicchio et al.,
2021}; \citeproc{ref-Gerosa:2019dbe}{Gerosa et al., 2019};
\citeproc{ref-Toubiana:2022vpp}{Toubiana et al., 2022}), and, crucially,
massive black hole binaries (MBHBs) with masses in the range of
\(10^4\)-\(10^7 M_\odot\) (\citeproc{ref-Klein:2015hvg}{Klein \& others,
2016}).

The presence of such a large number of signals requires joint inference
of all parameters across all sources present in the data, an approach
referred to as ``global fit'' (\citeproc{ref-Deng:2025wgk}{Deng et al.,
2025}; \citeproc{ref-Katz:2024oqg}{Katz et al., 2025};
\citeproc{ref-Littenberg:2023xpl}{Littenberg \& Cornish, 2023};
\citeproc{ref-Strub:2024kbe}{Strub et al., 2024}). This requires
evaluating hundreds of thousands of waveform templates, which can be
computationally expensive. To address this challenge, we present
\texttt{Phentax}, a \texttt{JAX} (\citeproc{ref-jax2018github}{Bradbury
et al., 2018}) re-implementation of the \texttt{IMRPhenomTHM} waveform
model (\citeproc{ref-Estelles:2020twz}{Estellés, Husa, et al., 2022}).
This is a state-of-the-art time-domain phenomenological waveform model
for comparable-mass binary black holes that includes the effects of
higher-order modes. Given this, \texttt{IMRPhenomTHM} is one of the
models currently used to simulate and analyze MBHB signals in LISA. With
\texttt{Phentax}, we provide a fast, easy-to-batch, and differentiable
implementation of this waveform model to accelerate the development of
LISA global fit algorithms.

\section{Statement of need}\label{statement-of-need}

Current global fit prototypes rely, at their core, on parallel-tempered
ensemble Markov Chain Monte Carlo (MCMC) samplers, in which a
(potentially large) number of MCMC chains are evolved in parallel. An
example is the \texttt{Eryn} sampler
(\citeproc{ref-Karnesis:2023ras}{Karnesis et al., 2023}) used in
(\citeproc{ref-Katz:2024oqg}{Katz et al., 2025}). These algorithms
require evaluating multiple likelihoods per iteration, which in turn
requires evaluating multiple waveform templates. It is therefore highly
beneficial to compute these likelihoods and templates in parallel rather
than sequentially. As a consequence, Graphical Processing Units (GPUs)
have been widely used in the LISA data analysis community to leverage
their parallel computing capabilities. \texttt{Phentax}, implemented in
\texttt{JAX}, can be easily run on GPUs, is designed to be batched,
i.e., to compute multiple waveforms in parallel, and provides
straightforward access to waveform derivatives. This can be particularly
useful for gradient-based samplers, such as the No-U-Turn sampler (NUTS)
(\citeproc{ref-2011arXiv1111.4246H}{Hoffman \& Gelman, 2011}) and for
the computation of Fisher information matrices
(\citeproc{ref-10.1098ux2frsta.1922.0009}{Fisher, 1922}).

Additionally, the non-stationary nature of the LISA power spectral
density (PSD) (\citeproc{ref-2020PhRvD.102h4062E}{Edwards et al., 2020})
makes the time-frequency domain a natural choice for analyzing LISA
data. This is also the natural domain for broadband, slowly-chirping
signals and thus holds significant potential for accelerating the
computation of the pre-merger phase in massive-black-hole binary merger
signals (\citeproc{ref-2025PhRvD.111j4044T}{Rodrigo \& Gerosa, 2025}).
In this context, \texttt{Phentax} provides direct access to the
time-frequency representation of the waveform, computed via the methods
presented in (\citeproc{ref-Bandopadhyay:2025fyx}{Bandopadhyay et al.,
2025}).

\section{State of the field}\label{state-of-the-field}

\texttt{Phentax} has been developed in the context of the analysis of
the ``Mojito Light'\,' common dataset (CD1L) produced by the LISA
Distributed Data Processing Center (DDPC). This is a simulated dataset
that includes, together with other LISA sources, 20 MBHB signals. These
have been generated using the \texttt{IMRPhenomTHM} waveform model
through the \texttt{phenomxpy} package
(\citeproc{ref-Garcia-Quiros:2025usi}{Garc{\'i}a-Quir{\'o}s et al., 2025}).
Before the release of CD1L, most LISA data analysis work relied on
frequency-domain MBHB waveforms, such as the \texttt{PhenomD/HM} models
(\citeproc{ref-Khan:2015jqa}{Khan et al., 2016};
\citeproc{ref-London:2017bcn}{London et al., 2018}). These enable
efficient computation of the waveform and the frequency-domain LISA
response (\citeproc{ref-Marsat:2018oam}{Marsat \& Baker, 2018}), but
they come at the expense of accuracy. In the future, different
time-domain waveform models will be used to generate MBHB signals in the
DDPC common datasets. In particular, for spin-precessing binaries, the
\texttt{IMRPhenomTPHM} (\citeproc{ref-Estelles:2021gvs}{Estellés,
Colleoni, et al., 2022}) and \texttt{SEOBNRv5PHM}
(\citeproc{ref-Ramos-Buades:2023ehm}{Ramos-Buades et al., 2023}) models
are expected to be used.

\texttt{Phentax} has been developed using \texttt{phenomxpy} as a guide.
We still decided to develop a separate package, rather than contribute
to \texttt{phenomxpy} because the original implementation is
fundamentally incompatible with \texttt{JAX}. In particular, the
implementation contained in \texttt{phenomxpy} does not allow for
vectorization and differentiation of the waveform computation, which are
crucial benefits of \texttt{Phentax}. At multiple points during the
development of this code, we have verified the correctness of the
waveform implementation against the original implementation in
\texttt{phenomxpy}. \texttt{Phentax} is not meant to provide a new
waveform model, but rather a tool tailored to the needs of global fit
prototypes, which largely benefit from the speedup this implementation
provides. The code itself remains applicable to stellar-mass black hole
binaries and can be used in the context of ground-based detectors as
well.

\section{Software design}\label{software-design}

\texttt{Phentax} heavily relies on \texttt{JAX}'s just-in-time
compilation (\texttt{jax.jit}) and vectorization (\texttt{jax.vmap})
capabilities. In particular, \texttt{Phentax} internally uses
\texttt{jax.vmap} to vectorize the waveform computation over a batch of
\(N_{\mathrm{batch}}\) input parameters, meaning that the user can
provide either floats or arrays for each waveform parameter.

The main entry point of the code is the \texttt{IMRPhenomTHM} class,
contained in the \texttt{phentax.waveform} module. This class provides a
simple interface to compute the waveform in the time domain, and it
gives access to any waveform component the user may be interested in:
amplitude and phase of each mode, strain components \(h_{lm}\) with and
without the contribution of spherical harmonics \(Y_{lm}\), and the plus
and cross polarizations \(h_{+,\times}\).

In terms of mode content, this implementation includes the
\((l, \,m) = (2, \pm 2)\), \((2, \pm 1)\), \((3,\pm 3)\),
\((4, \pm 4)\), \((5, \pm 5)\) modes, where the negative \(m\) modes are
computed using the symmetry relation
\(h_{l,-m} = (-1)^l \, h_{lm}^\dagger\). This is possible because the
model implemented here is restricted to non-precessing binaries, for
which the symmetry relation holds. The user can choose to compute the
waveform with all modes or only a subset of them.

As \texttt{Phentax} relies on \texttt{jax.jit}, \texttt{JAX} will trace
and compile the waveform computation the first time it is called. This
introduces a significant initial overhead, which we refer to as the
``warm-up'' time, and it is therefore crucial to minimize the number of
times the waveform has to be traced and compiled. In a jitted function,
recompilation is triggered as one of the inputs changes shape or type.
This led us to design the \texttt{IMRPhenomTHM} class to be initialized
with an argument that sets the maximum observation time \(T\), which is
used to preallocate the time and waveform arrays. This is necessary when
computing the waveform from a minimum frequency \(f_{\rm min}\) onwards,
as systems with different masses and the same initial frequency will
take different times to reach the merger. After the arrays are
preallocated with shape \((N_{\mathrm{batch}}, \, N_{\mathrm{times}})\),
for each entry in the batch the waveform is computed and stored in the
matching row of the preallocated arrays. The unused parts of the arrays
are set to zero for the waveform quantities, and to the last valid time
for the time arrays. Along with the time arrays, the class methods also
return a boolean mask that identifies the valid part of the waveform,
i.e., the part that is not padded with zeros.

Alternatively, instead of a densely sampled time array, the waveform can
be computed on an adaptive time array generated internally from the
leading-order phase evolution of the \((2,2)\) mode. We refer the reader
to the documentation for more details.

\begin{figure}[h!]
 \centering
 \includegraphics[width=0.9\linewidth]{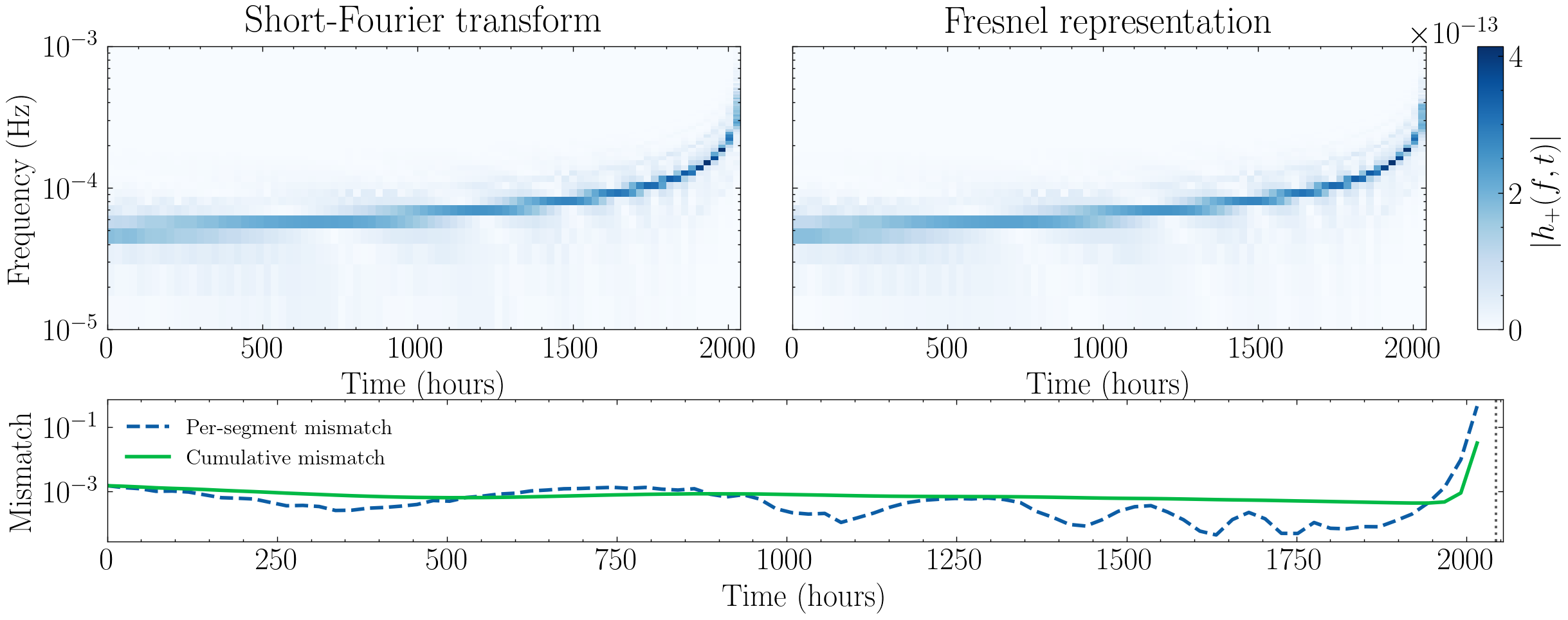}
 \caption{Comparison of the $h_+$ polarization in the time-frequency domain, between the analytic Fresnel-based time-frequency representation and a short-Fourier transform computed from $h_+(t)$. Runtime cost for a signal of this length: $\sim 150$ microseconds.}
 \label{fig:tf}
\end{figure}

A similar design philosophy is adopted for the time-frequency
implementation of \texttt{Phentax}. This also relies on tracing and
compilation via \texttt{jax.jit} against a constant time grid over which
the waveform must be evaluated, and a batch size of
\(N_{\mathrm{batch}}\); changing either of these two parameters triggers
recompilation. In this framework, we circumvent instantiating the dense
time-domain waveform arrays, instead using the source parameters to
directly evaluate amplitudes and phases on the much sparser
time-frequency grid. This significantly reduces the computational cost
of generating long waveforms corresponding to binaries with low total
mass; see e.g.~Fig. \ref{fig:tf}. For further details on the
time-frequency formalism adopted here, see
(\citeproc{ref-Bandopadhyay:2025fyx}{Bandopadhyay et al., 2025}).

\begin{figure}[h!]
 \centering
 \begin{subfigure}{0.45\textwidth}
 \centering
 \includegraphics[width=\linewidth]{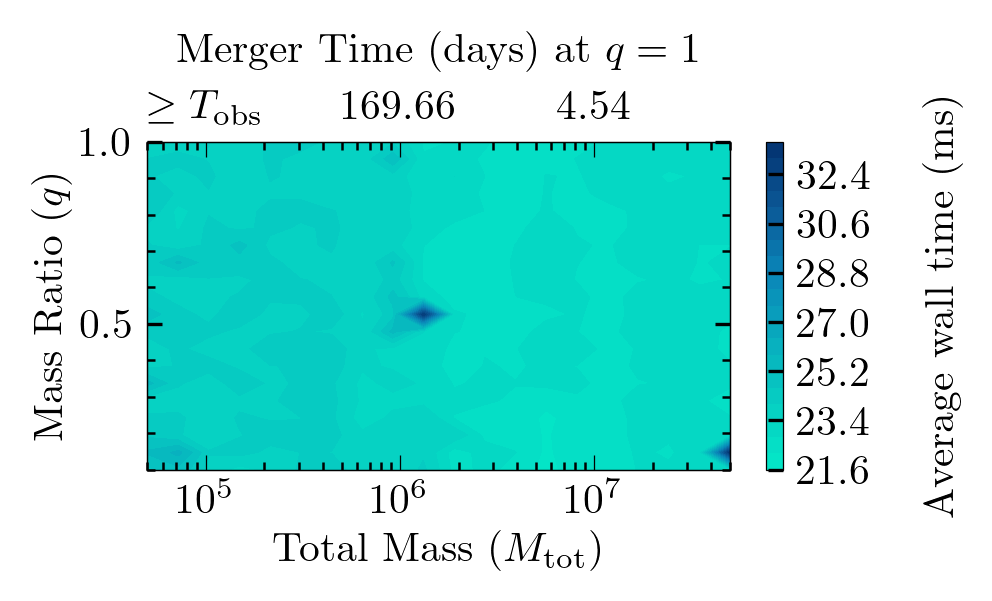}
 \caption{GPU timings}
 \end{subfigure}
 \begin{subfigure}{0.45\textwidth}
 \centering
 \includegraphics[width=\linewidth]{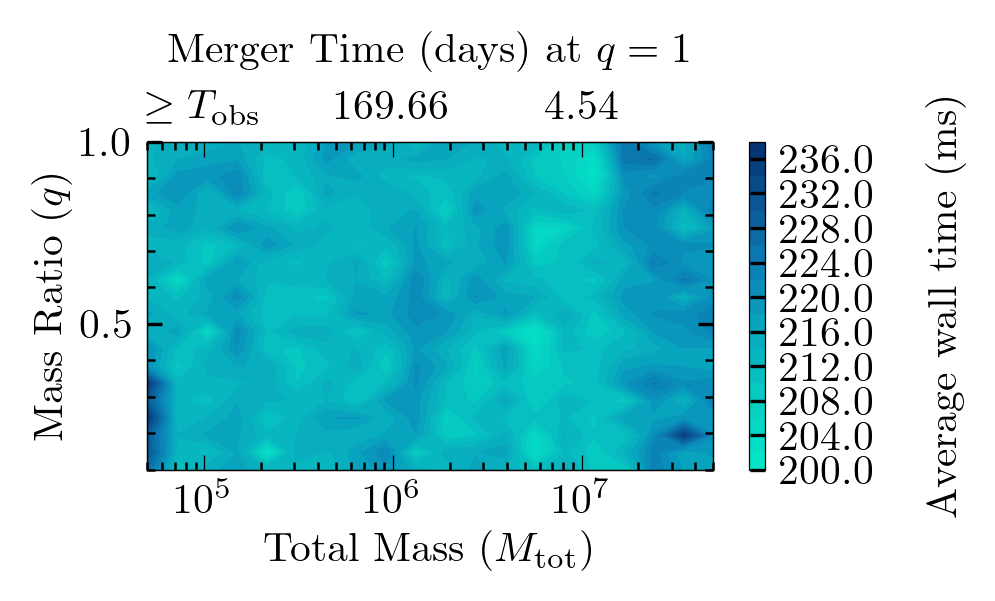}
 \caption{CPU timings}
 \end{subfigure}
 \caption{Average wall-clock time to compute a single waveform (after warm-up) as a function of the total mass $M_{\rm tot}$ and mass ratio $q$. The remaining parameters are held fixed to $\{s_{1z}, s_{2z}, d_{\rm L}, \iota, \phi_{\rm ref}, \psi, f_{\rm min}, \mathrm{d}t\} = \{0.9, \, 0.3, \, 500 \, \mathrm{Mpc}, \, \pi / 3, \, 0.0, \, 1.0,\,  9\times 10^{-5} \, \mathrm{Hz}, 10 \, \mathrm{s}\}$. Here, the observation time is fixed to 6 months, and the averages are computed over $50$ realizations.
 For each value of $M_{\rm tot}$, the top x-axis shows the corresponding time to merger for equal-mass systems.
 }
 \label{fig:timings_Mt_q}
\end{figure}

\begin{figure}[h!]
 \centering
 \begin{subfigure}{0.45\textwidth}
 \centering
 \includegraphics[width=\linewidth]{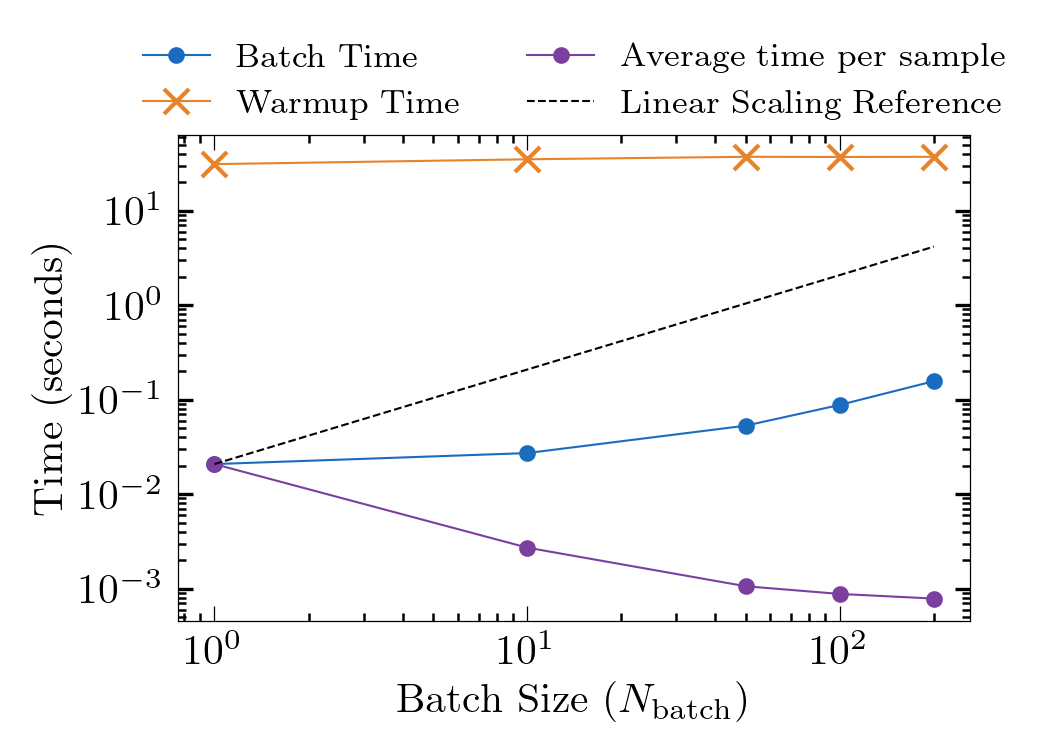}
 \caption{GPU timings}
 \end{subfigure}
 \begin{subfigure}{0.45\textwidth}
 \centering
 \includegraphics[width=\linewidth]{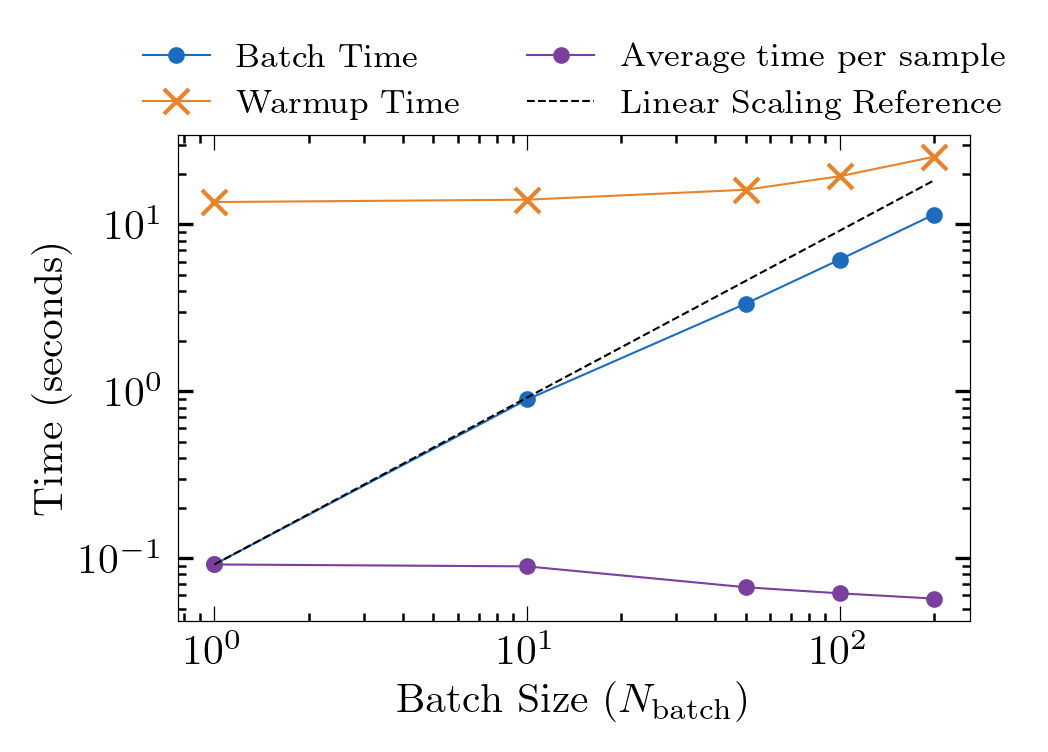}
 \caption{CPU timings}
 \end{subfigure}
 \caption{Wall-clock time to compute a batch of waveforms (blue) and the corresponding time for a single waveform (purple) as a function of the batch size. Here, the observation time is fixed to 2 months. As a reference, we also report the warm-up time for each batch size (orange), and the linear scaling (black dashed line) reference $t(n=N_{\rm batch}) = N_{\rm batch} \, t(n=1)$. We use $t(n)$ to denote the elapsed time required to generate a batch of $n$ waveforms. As expected, the GPU can fully leverage the parallelization capabilities of the hardware, while the CPU batch timing remains consistent with a linear scaling. Each waveform of a batch is generated with parameters randomly drawn around the ones used in Fig. \ref{fig:timings_Mt_q}, with mean total mass and mass ratio set to $M_{\rm tot} = 10^6 M_\odot$ and $q = 0.7$, respectively.
 }
 \label{fig:timings_batch_size}
\end{figure}

We have designed \texttt{Phentax} with a focus on GPU performance, but
it can also be run on CPUs. We report timings on both architectures in
Fig. \ref{fig:timings_Mt_q} and Fig. \ref{fig:timings_batch_size}.

\section{Research impact statement}\label{research-impact-statement}

At the time of writing, \texttt{Phentax} is being heavily used in the
analysis of the Mojito Light common dataset (CD1L) produced by the LISA
Distributed Data Processing Center (DDPC). In particular, the software
has been adopted by global fit pipelines in both the DDPC and the NASA
Science Ground Segment (NSGS) to produce MBHB templates. Publications of
the results of these analyses are expected in the near future.

Moreover, \texttt{Phentax} has also been adopted in searches and
early-alert pipelines developed with CD1L as the target. There, the goal
is to produce early alerts of MBHB mergers to enable multi-messenger
follow-up observations. Therefore, many pipelines in this context
leverage Simulation-Based Inference (SBI) techniques to enable fast
inference of the source parameters on the fly. This is only possible
after an expensive training phase that requires generating a large
number of waveform templates. The code's speed and batching capabilities
can make the training phase significantly cheaper, which is crucial when
pipelines are still in their early development stage, and training may
need to be repeated (in principle) multiple times.

The efficient time-frequency representation available within
\texttt{Phentax} is being used to prototype other pre-merger
alert/analysis scenarios. This representation is particularly well
suited to fast evaluation of waveforms from low-mass binaries
\((\sim 10^5 - 10^6 \, \rm{M_{\odot}})\), which are present within the
LISA detector frequency band for extended periods of time. This approach
is currently being applied to developing pre-merger search strategies
for Mojito Light.

\section{Availability}\label{availability}

\texttt{Phentax} is available via
\href{https://pypi.org/project/phentax/}{PyPI}. The code is hosted on
\href{https://github.com/asantini29/phentax}{GitHub}, and the
documentation is available at
\url{https://asantini29.github.io/phentax}. The latter includes simple
tutorials on how to get started with the package.

\section{AI usage disclosure}\label{ai-usage-disclosure}

We acknowledge the usage of generative AI tools for the development of
this work. Anthropic's \texttt{Claude\ Code} and Google's
\texttt{Gemini} were used to produce part of the unit tests, debug the
\texttt{JAX} implementation, produce parts of the documentation, and
refactor parts of the codebase at different stages of the project. We
have reviewed and edited the content to ensure consistency with the
original intended meaning and to preserve the integrity of the work.
\texttt{Grammarly} has been used to improve the text's grammar and
clarity.

\section{Acknowledgements}\label{acknowledgements}

We thank Jules Perret, Senwen Deng, Solano Felicio, and Jonah George for
useful comments and bug reports on the repository. We also acknowledge
the members of the Distributed Data Processing Center (DDPC) for
feedback and comments. A.S. gratefully acknowledges support from the
German Space Agency (DLR) with funding of the Bundesministerium für
Wirtschaft und Klimaschutz with a decision of the Deutsche Bundestag
(Project Ref. No.~FKZ 50 OQ 2301). Timings and benchmarks reported here
were performed on the \texttt{Saraswati} GPU cluster at the Max Planck
Institute for Gravitational Physics in Potsdam, using NVIDIA A100 GPUs.

\section*{References}\label{references}
\addcontentsline{toc}{section}{References}

\protect\phantomsection\label{refs}

\end{document}